\documentclass[letterpaper, 10pt, compsoc]{IEEEtran}

\usepackage{graphicx}
\usepackage[caption=false]{subfig}
\usepackage{multirow}
\usepackage{array}
\usepackage{epsfig} 
\DeclareGraphicsExtensions{.pdf,.eps,.png,.jpg,.mps}
\usepackage{minibox}

\usepackage{amsmath} 
\usepackage{amssymb}  
\usepackage{algorithm}
\usepackage{algpseudocode}
\usepackage{cite}
\usepackage{hyperref}
\usepackage{breakurl}
\usepackage{footnote}
\usepackage{mathtools}
\newtheorem{theorem}{Theorem}

\usepackage{flushend}
\usepackage{todonotes}
\usepackage{fmtcount}

\allowdisplaybreaks
\begin{document}

\title{Energy-Efficient Scheduling for Homogeneous Multiprocessor Systems} 
\author{Mason Thammawichai,~\IEEEmembership{Student Member,~IEEE}
and Eric C. Kerrigan,~\IEEEmembership{Member,~IEEE}

\IEEEcompsocitemizethanks{\IEEEcompsocthanksitem Mason Thammawichai is with the Department of Aeronautics, Imperial College London, London SW7 2AZ, UK \protect\\
E-mail: m.thammawichai12@imperial.ac.uk
\IEEEcompsocthanksitem Eric C.~Kerrigan is with the Department of Electrical \& Electronic Engineering  and the Department of Aeronautics, Imperial College London, London SW7 2AZ, UK. \protect\\
E-mail: e.kerrigan@imperial.ac.uk}}
\IEEEtitleabstractindextext{%
\begin{abstract}
We present a number of novel algorithms, based on mathematical optimization formulations, in order to solve a homogeneous  multiprocessor scheduling problem, while minimizing the total energy consumption. In particular, for a system with a discrete speed set, we  propose solving a tractable linear program. Our formulations are based on a fluid model and a global scheduling scheme, i.e.\ tasks are allowed to migrate between processors. The new methods are compared with three global energy/feasibility optimal workload allocation formulations. Simulation results illustrate that our methods achieve both feasibility and energy optimality and outperform existing methods for constrained deadline tasksets. Specifically, the results provided by our algorithm can achieve up to an 80\% saving compared to an algorithm without a frequency scaling scheme and up to 70\% saving compared to a constant frequency scaling scheme for some simulated tasksets. Another benefit is that our algorithms can solve the scheduling problem in one step instead of using a recursive scheme. Moreover, our formulations can solve a more general class of scheduling problems, i.e.\ any periodic real-time taskset with arbitrary deadline. Lastly, our algorithms can be applied to both online and offline scheduling schemes.
\end{abstract}}

\maketitle

\IEEEraisesectionheading{\section{Introduction}}
\IEEEPARstart{D}ue to higher computational power demands in modern computing systems, e.g. sensor networks, satellites, multi-robot systems, as well as personal embedded electronic devices, a system's power consumption is significantly increase. Therefore, an efficient energy management protocol is required in order to balance the power consumption and workload requirement of a system. With this motivation, this paper is aimed at using an optimization-based approach to develop an algorithm for a multiprocessor real-time scheduling problem with the goal of minimizing energy consumption. 

The dynamic power consumption of a CMOS processor is often represented as a function of both  clock frequency and supply voltage~\cite{Rabaey:02}. Thus, a dynamic voltage and frequency scaling (DVFS) scheme is often applied to reduce processor power consumption. DVFS is now commonly implemented in various computing systems, both at hardware and software levels. Examples include Intel's SpeedStep technology, AMD's PowerNow! technology and the Linux Kernel Power Management Scheme.

\subsection{Terminologies and Definitions}\label{sec:def} 
This section provides basic terminologies and definitions used throughout the paper.

\indent\textbf{Speed $s$}: The speed $s$ of a processor is defined as the ratio between the operating frequency $f$ and the maximum system frequency $f_{max}$, i.e.\ $s:=f/f_{max}$. We denote with $s_{min}$ the minimum execution speed of a processor.

\indent\textbf{Task $T_i$}: A task $T_i$ is defined as a triple $T_i:=(c_i,d_i,p_i)$; $c_i$ is the required number of processor cycles, $d_i$ is the task's deadline and $p_i$ is the task's period. If the task's deadline is equal to its period, the task is said to have an `implicit deadline'. The task is considered to have a `constrained deadline' if its deadline is not larger than its period, i.e. $d_i \leq p_i$. In the case that the task's deadline can be less than, equal to, or greater than its period, it is said to have an `arbitrary deadline'.

\indent\textbf{Job $T_{ij}$}: A job $T_{ij}$ is defined as the $j^{th}$ instance of task $T_i$, where $j\geq 1$. $T_{ij}$ arrives at time $(j-1)p_i$, has the required execution cycles $c_i$ and a deadline at time $(j-1)p_i+d_i$. 
 
\indent\textbf{Taskset}: The taskset is defined as a set composed of real-time tasks.

\indent\textbf{Minimum Execution Time $\underline{x}_i$}: The minimum execution time $\underline{x}_i$ of a task $T_i$ is the execution time of the task when the task is executed at $f_{max}$,~$\underline{x}_i:=c_i/f_{max}$.

\indent\textbf{Task Density $\delta_i(s_i)$}: The task density $\delta_i(s_i)$ of a task $T_i$ executed at a speed $s_i$ is defined as the ratio between the task execution time and the minimum of its deadline and its period, i.e.\ $\delta_i(s_i):=c_i/(s_if_{max}\min\{d_i,p_i\})$. When all tasks are assumed to have an implicit deadline, this is often called `task utilization'.

\indent\textbf{Taskset Density $D(s_i)$}: The taskset density $D(s_i)$ is defined as the summation of all task densities in the taskset, i.e.~$D(s_i):=\sum_{i=1}^n \delta_i(s_i)$, where $n$ is a number of tasks in the taskset. The minimum taskset density $D$ is given by $D:=\sum_{i=1}^n \delta_i(1)$.

\indent\textbf{Feasibility Optimal}: Feasibility optimality of a homogeneous multiprocessor system is obtained when the upper bound on the minimum taskset density $D$, for which the algorithm is able to construct a valid schedule such that no deadlines are missed, equals the total number of processors in the system.

\indent\textbf{Scheduling Scheme}: Multiprocessor scheduling can be classified according to the task migration scheme\footnote{A task \textit{migration} occurs when a task execution is suspended on one processor and continues on another processor, while \textit{pre-emption} is used when the execution of a task on a processor is suspended in order to start executing another task.}: a `global scheduling scheme'  allows task migration between processors and a `partitioned scheduling scheme' does not allow task migration.

\subsection{Related Work} 
\begin{figure}[t]
\centering
\includegraphics[width=0.5\textwidth]{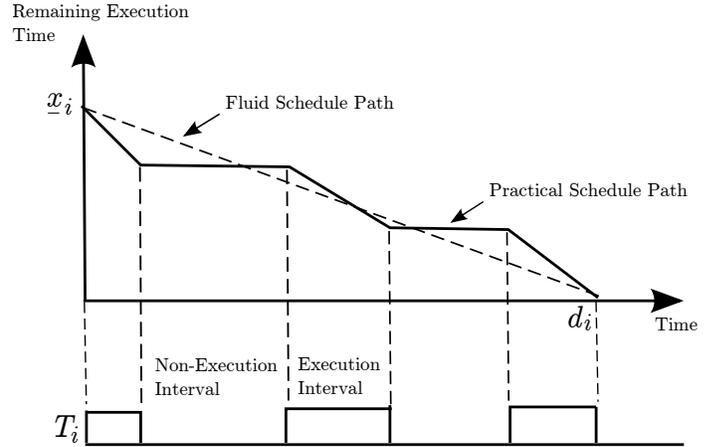}
\caption{Fluid Schedule and a Practical Schedule \cite{cho06}}
\label{fig:fmodel}
\end{figure}

There are at least two well-known feasibility optimal homogeneous multiprocessor scheduling algorithms of an implicit deadline taskset that are based on a fluid scheduling model: Proportionate-fair (Pfair)~\cite{baruah96} and Largest Local Remaining Execution Time First (LLREF)~\cite{cho06}. Both Pfair and LLREF are global scheduling algorithms. A fluid model, shown in Figure~\ref{fig:fmodel}, is the ideal schedule path of a task $T_i$, where the remaining minimum execution time is represented by a straight line and the slope of the line is the task execution speed $s_i$. 

By introducing the notion of \emph{fairness}, Pfair ensures that at any instant no task is one or more quanta (time intervals) away from the task's fluid path. However, the Pfair algorithm suffers from a significant run-time overhead, because tasks are split into several segments, incurring frequent algorithm invocations and task migrations. 

To overcome these disadvantages, the LLREF algorithm pre-empts a task at two scheduling events within each time interval~\cite{cho06}. One occurs when the remaining time of an executing task is zero and the other event happens when the difference between a deadline and a remaining execution time of a task is zero. 

By incorporating a DVFS scheme with a fluid model,~\cite{funaoka108} proposed the real-time static voltage and frequency scaling (RT-SVFS) algorithm, which allows the slope of a fluid schedule to vary between 0 and 1. To improve the performance of the open-loop static algorithm of~\cite{funaoka108}, a closed-loop dynamic algorithm is proposed in~\cite{funaoka208}. Their dynamic algorithm considered uncertainty in the task execution time. By extending~\cite{funaoka108}, an energy-efficient scheduling algorithm for real-time tasks with unknown arrival time, known as a sporadic real-time task, was proposed in~\cite{zhang11}. 

Deadline Partitioning (DP)~\cite{levin2010} is the technique that partitions time into intervals bounded by two successive task deadlines, after which each task is allocated the workload and is scheduled at each time interval. A simple optimal scheduling algorithm, called DP-WRAP, was presented in~\cite{levin2010}. The DP-WRAP algorithm partitions time according to the DP technique and, at each time interval, the tasks are scheduled using McNaughton's wrap around algorithm~\cite{mc1959}. McNaughton's wrap around algorithm aligns all task workloads along a number line, starting at zero, then splits tasks into chunks of length 1 and assigns each chunk to the same processor.

However, the algorithms that are based on the fairness notion~\cite{cho06,funaoka108,funaoka208,levin2010,funk2012,fei2012} are schedulability optimal, but have hardly been applied in a real system, since they suffer from high scheduling overheads, i.e.~task pre-emptions and migrations. Recently, two schedulability optimal algorithms that are not based on the notion of fairness have been proposed. One is the RUN algorithm~\cite{regnier2011}, which uses a dualization technique to reduce the multiprocessor scheduling problem to a series of uniprocessor scheduling problems. The other is U-EDF~\cite{nelissen2012}, which generalises the earliest deadline first (EDF) algorithm to multiprocessors by reducing the problem to EDF on a uniprocessor.

Alternatively to the above methods, the multiprocessor scheduling problem can also be formulated as an optimization problem. However, since the problem is NP-hard, in general~\cite{lawler1983},  polynomial-time heuristic methods are often used. An example of these approaches can be found in~\cite{aydin2003}, which addresses the energy-aware multiprocessor partitioning problem and \cite{hoon2014}, which proposes an energy and feasibility optimal global framework for a two-type multicore platform. In general, the tasks are partitioned among the set of processors, followed with computing the running frequency. Among all of the feasibility assignments, an optimal energy consumption assignment is chosen by solving a mathematical optimization problem, where the objective is to minimize energy. The constraints are to ensure that all tasks will meet their deadlines and only one processor is assigned to a task. In partitioned scheduling algorithms, such as~\cite{aydin2003}, once a task is assigned to a specific processor, the multiprocessor scheduling problem is reduced to a set of uniprocessor scheduling problems, which is well studied~\cite{chen2007}. However, a partitioned scheduling method cannot provide an optimal schedule. 

\subsection{Contribution}
In this paper, we propose three mathematical optimization problems to solve a periodic hard real-time task scheduling problem on homogeneous multiprocessor systems with DVFS capabilities. First is an MINLP, which adopts the fluid model used in~\cite{baruah96, cho06, funaoka108} to represent a scheduling dynamic. The MINLP relies on the optimal control of a suitably-defined dynamic to globally solve for a valid schedule, while solutions are obtained by solving each instance of the time interval obtained using the DP technique~\cite{cho06, funaoka108, levin2010}. By determining a fraction of task's execution time and operating speed, rather than task assignments, the same scheduling problem can be formulated as an NLP. Lastly, we propose an LP for a system with discrete speed levels. Our work presents homogeneous multiprocessor scheduling algorithms that are both feasibility optimal and energy optimal. Furthermore, our formulations are capable of solving any periodic tasksets, i.e.~implict, constrained and arbitary deadlines.

\subsection{Outline of Paper}
This paper is organized as follows: Section~\ref{sec:prob} defines our scheduling problem in detail. Three mathematical optimization formulations to solve the same multiprocessor scheduling problem are proposed in Section~\ref{sec:soln}. The simulation setup and results are presented in Section~\ref{sec:sim}. Finally, conclusions and future work are discussed in Section~\ref{sec:concl}.

\section{Problem Formulation} \label{sec:prob}
\subsection{Task and Processor Model} \label{sec:taskmodel}
We consider a set of $n$ periodic real-time tasks that are to be partitioned on $m$ identical processors, where each processor's voltage/speed can be adjusted individually. All tasks are assumed to start at the same time. The hyperperiod $L$ is defined as the least common multiple of all task periods. The tasks can be pre-empted at any time, do not share resources and do not have any precedence constraints. It is assumed that $D \leq m$ in order to guarantee the existence of a valid schedule.

Below, we will refer to the sets $I:=\{1,\ldots,n\}$, $J_i:=\{1,\ldots,L/p_i\}$ and $K:=\{1,\ldots,m\}$. The remaining minimum execution time of job $T_{ij}$ at time $t$ will be denoted by  $x_{ij}(t)$.   Note that $\forall i,\forall j,\forall k,\forall t$ will be used as short-hand  for $\forall i\in I,\forall j\in J_i,\forall k\in K,\forall t\in [0,L]$, respectively.

\subsection{Energy Consumption Model} \label{sec:gmodel}
For CMOS processors, the total power consumption is often simply expressed as an increasing function of the form 
\begin{equation}
P(s):=P_d(s)+P_s=\alpha s^\beta+P_s, \label{eqn:Pd_lit}
\end{equation} 
\noindent where $P_d(s)$ is the dynamic power consumption due to the charging and discharging of CMOS gates, $\alpha >0$  and $\beta \geq 1$ are hardware-dependent constants and the static power consumption $P_s$, which is mostly due to leakage current, is assumed to be either constant or zero \cite{Rabaey:02}.

The total energy consumption from executing a task can be expressed as a summation of the active and idle energy consumed, i.e.\ $E = E_{active} + E_{idle}$, where $E_{active}$ is the energy consumed when the processor is busy executing the task and $E_{idle}$ is the energy consumed when the processor is idle. The energy consumed by executing and completing a task at a constant speed $s$ is
\begin{subequations} 
\begin{align}
E(s) &= E_{active}(s) + E_{idle}(s) \\
&=\frac{c}{f_{max}} \frac{(P_{active}(s)-P_{idle})}{s} + P_{idle}d  \\
&=\frac{x(P_{active}(s)-P_{idle})}{s} + P_{idle}d,
\end{align}
\end{subequations} 
 
\noindent where $P_{active}(s):=P_{d}(s)+P_{s}$ is the  power consumption in the active interval, $P_{idle}$ is the power consumption during the idle period. $P_{idle}>0$ and $P_{s}>0$ will be assumed to be constants. Note that $P_{active}(s)-P_{idle}$ is strictly greater than zero and that the last term $P_{idle}d$ is not a function of the speed.  

\subsection{Scheduling as an Optimal Control Problem}\label{sec:minlp}
The objective is to minimize the total energy consumption of executing a periodic taskset within the hyperperiod $L$.
The scheduling problem  can therefore be formulated as the following infinite-dimensional continuous-time optimal control problem: 
\begin{subequations} \label{cprob}
\begin{align}
	&
	\underset{\begin{subarray}{c}
           x_{ij}(\cdot),a_{ijk}(\cdot),\\
	  s_{ijk}(\cdot),i\in I,\\
	j\in J_i,k\in K \end{subarray}}
	{\text{minimize}} \
	\mathrlap{
	\sum_{i,j,k}
	\int_0^L
	\ell(a_{ijk}(t),s_{ijk}(t))dt}  \\
	 &\text{subject to }  &\nonumber\\
	     &x_{ij}((j-1)p_i) = \underline{x}_i,&&\forall i,j\label{ho1}\\ 
			 &x_{ij}((j-1)p_i+d_i) = 0,&&\forall i,j \label{ho2}\\
       &\dot{x}_{ij}(t) =-\sum_{k} s_{ijk}(t)a_{ijk}(t),&&\forall i,j,t, \text{ a.e.} \label{ho3}\\       
			 &\sum_{k} a_{ijk}(t) \leq 1,&&\forall i,j,t \label{ho4}\\
			 &\sum_{i,j} a_{ijk}(t)\leq 1,&&\forall k,t \label{ho5}\\
       &s_{min} \leq s_{ijk}(t) \leq 1,&&\forall i,j,k,t \label{ho6}\\
			 &a_{ijk}(t) \in \{0,1\},&&\forall i,j,k,t \label{ho7}
\end{align}
\end{subequations}
where 
\begin{align*}
\ell(a,s):=a(P_{active}(s)-P_{idle}), 
\end{align*}
 $s_{ijk}(t)$ is the execution speed of job $T_{ij}$ at time $t$ and $a_{ijk}(t)$ is used to indicate processor assignment, i.e.\ $a_{ijk}(t) = 1$ if and only if job $T_{ij}$ is active on processor $k$ at time $t$. 

The initial conditions on the minimum execution time of all jobs are specified in~(\ref{ho1}) and job deadline constraints are specified by~(\ref{ho2}). The fluid model of the scheduling  dynamic is given in~(\ref{ho3}), where the state is $x$ and the control inputs  are   $a$ and $s$. The constraint that each job is assigned to at most one processor at a time is ensured by~(\ref{ho4}) and~(\ref{ho5}) enforces that each processor is assigned to at most one job at a time. Upper and lower bounds on the processor speed are given in~(\ref{ho4}).  The binary nature of job assignment variables are given by~(\ref{ho7}).

\section{Solving the Scheduling Problem with Finite-dimensional Mathematical Optimization} \label{sec:soln}
This section provides details on three mathematical optimization problems to solve the same scheduling problem defined in Section~\ref{sec:minlp}. The original problem~(\ref{cprob}) will be discretized by introducing piecewise constant constraints on the control inputs $s$ and~$a$.

Let $\{\tau_0, \tau_1, \ldots, \tau_{N}\}$,  which we will refer to as the major grid,  be the set of  time instances corresponding to the distinct arrival times and deadlines of all jobs within the hyperperiod~$L$, where $0=\tau_0\leq \tau_1  \leq \cdots \leq \tau_{N} = L$.

\subsection{Mixed-integer Nonlinear Program (MINLP-DVFS)}\label{sec:dis1}
The above scheduling problem, subject to piecewise constant constraints on the control inputs, is most naturally formulated as an MINLP, as defined below.

Though it has been observed by~\cite{cho06} that it is often the case that tasks might have to be split in order to make the entire taskset schedulable, the runtime overheads caused by context switches can jeopardize the performance. Therefore, a variable discretization time step~\cite{gerdts2006} method is applied in a so-called minor grid, so that the solution to our scheduling problem does not depend on the size of the discretization time step. Let $\{\tau_{\mu,0},\ldots,\tau_{\mu,M}\}$ denote the set of time instances on a minor grid within the time interval $[\tau_\mu,\tau_{\mu+1}]$ with $\tau_\mu=\tau_{\mu,0}\leq \cdots \leq \tau_{\mu,M}=\tau_{\mu+1}$, so that  $\{\tau_{\mu,1},\ldots,\tau_{\mu,M-1}\}$ is to be determined for all $\mu$ from solving an appropriately-defined optimization problem.

Let $\forall \mu$ and $\forall \nu$ be short-hand  for $\forall \mu\in U:=\{0,\ldots,N-1\}$ and $\forall \nu\in V:=\{0,\ldots,M-1\}$. Define the notation $[\mu,\nu]:= (\tau_{\mu,\nu}),\forall \mu,\nu$  and the discretized state and input sequences as
\begin{subequations}
\begin{align*}
&x_{ij}[\mu,\nu] := x_{ij}(\tau_{\mu,\nu}), && \forall i,j,\mu,\nu\\
&s_{ijk}[\mu,\nu] := s_{ijk}(\tau_{\mu,\nu}), &&  \forall i,j,k,\mu,\nu\\
&a_{ijk}[\mu,\nu] := a_{ijk}(\tau_{\mu,\nu}), &&  \forall i,j,k,\mu,\nu
\end{align*}
\end{subequations}

If $s_{ijk}(\cdot)$ and $a_{ijk}(\cdot)$ are  constant in-between time instances on the minor grid, i.e.\ 
\begin{subequations}
\label{eq:pwc}
\begin{align}
s_{ijk}(t) = s_{ijk}[\mu,\nu],\ &\forall t \in [\tau_{\mu,\nu},\tau_{\mu,\nu+1}),\mu,\nu\\
 a_{ijk}(t) = a_{ijk}[\mu,\nu],\ &\forall t \in [\tau_{\mu,\nu},\tau_{\mu,\nu+1}),\mu,\nu
\end{align}
\end{subequations}
 then it is easy to show that the solution of the scheduling dynamic~(\ref{ho3}) is given by
 \begin{subequations}
 \label{dprob}
\begin{multline}
x_{ij}[\mu,\nu+1]=x_{ij}[\mu,\nu]\\
-h[\mu,\nu]\sum_{k}s_{ijk}[\mu,\nu]a_{ijk}[\mu,\nu],
\quad \forall i,j,\mu,\nu\label{eq:dyn}
\end{multline}
where  \begin{align} h[\mu,\nu]:=\tau_{\mu,\nu+1}-\tau_{\mu,\nu},\forall \mu,\nu. \end{align}

Let $\Lambda$ denote the set of all jobs within hyperperiod $L$, i.e. $\Lambda:=\{T_{ij}|i\in I,j\in J_i\}$. Define a function $\Phi_A:\Lambda\rightarrow U\times V$ by $\Phi_A(T_{ij}):= (\mu,\nu)$ such that $\tau_{\mu,\nu}=(j-1)p_i,~\forall T_{ij}\in\Lambda$ and a function $\Phi_D:\Lambda\rightarrow U\times V$ by $\Phi_D(T_{ij}):=(\mu,\nu)$ such that $\tau_{\mu,\nu} = (j-1)p_i+d_i,~\forall T_{ij}\in\Lambda$. 

The original problem~(\ref{cprob}) subject to the piecewise constant constraints on the inputs~\eqref{eq:pwc}  is therefore equivalent to the following finite-dimensional MINLP:
\begin{align}
& \underset{\begin{subarray}{c}
       x_{ij}[\cdot],s_{ijk}[\cdot],\\
			 a_{ijk}[\cdot],h[\cdot],\\
			 i\in I,j\in J_i,k\in K\end{subarray}}
			{\text{minimize}} \mathrlap{\sum_{\mu,\nu,i,j,k}h[\mu,\nu]\ell(a_{ijk}[\mu,\nu],s_{ijk}[\mu,\nu])}\\
	& \text{subject to (\ref{eq:dyn}) and} \nonumber\\
	     &x_{ij}[\Phi_A(T_{ij})]=\underline{x}_i & &\forall i,j \\ 
			 &x_{ij}[\Phi_D(T_{ij})]= 0,  & &\forall i,j \\ 	 
			 &\sum_k a_{ijk}[\mu,\nu] \leq 1,&&\forall i,j,\mu,\nu\\ 
			 &\sum_{i,j} a_{ijk}[\mu,\nu]\leq 1,&&\forall k,\mu,\nu\\
       &s_{min} \leq s_{ijk}[\mu,\nu] \leq 1, &&\forall i,j, k,\mu,\nu\\ 
			 &a_{ijk}[\mu,\nu] \in \{0,1\}, &&\forall i,j, k,\mu,\nu \label{hd0}\\ 
			 &0 \leq h[\mu,\nu], &&\forall \mu,\nu \label{hd1}\\
			 &\sum_{\nu}h[\mu,\nu] = \tau_{\mu+1}-\tau_\mu, &&\forall \mu \label{hd2}
\end{align}
\end{subequations}

\noindent where 
 (\ref{hd1})--\eqref{hd2} enforce upper and lower bounds on the discretization time steps. 
 
 \begin{theorem}
 If a solution exists to~\eqref{ho1}--\eqref{ho7} and $M \geq \max\{n-1,m-1\}$, then a solution exists to problem~(\ref{dprob}).
 \end{theorem}
 \begin{IEEEproof}
 Follows from the fact that if a solution exists to~\eqref{ho1}--\eqref{ho7}, then the DP-WRAP scheduling algorithm can find a valid schedule with at most $n-1$ context switches and $m-1$ migrations per slice~\cite[Thm~6]{levin2010}.   
 \end{IEEEproof}

\subsection{Continuous Nonlinear Program (NLP-DVFS)}
This section proposes an NLP formulation without integer variables to solve the  problem in Section~\ref{sec:minlp}. The idea is to relax the binary constraints in~(\ref{hd0}) so that the value of $a$ can be interpreted as the fraction of a time interval during which the job is executed on a processor. 

Moreover, the number of variables can also be reduced compared to problem~\eqref{dprob}, since the processor assignment information does not help in finding the task execution order, hence $k$ is dropped from the subscripts in the notation. That is, partitioning time using only the major grid (i.e.\ $M=1$) is enough to find a feasible schedule if a solution exists to the original problem~\eqref{cprob}.  Since we only need a major grid, we define the notation $[\mu] :=(\tau_\mu)$.  

Consider now the following finite-dimensional NLP:
\begin{subequations} 
\label{prob2d}
\begin{align}
& \underset{\begin{subarray}{c}
       x_{ij}[\cdot],s_{ij}[\cdot],\\
			 a_{ij}[\cdot],i\in I,j\in J_i\end{subarray}}
			 {\text{minimize}} \quad\mathrlap{\sum_{\mu,i,j}(\tau_{\mu+1}-\tau_\mu) \ell(a_{ij}[\mu],s_{ij}[\mu])}\\
	& \text{subject to } \nonumber\\
	     &x_{ij}[\Phi_A(T_{ij})]=\underline{x}_i, & &\forall i,j \label{u1}\\ 
			 &x_{ij}[\Phi_D(T_{ij})]= 0,  & &\forall i,j \\ 
       &x_{ij}[\mu+1]=x_{ij}[\mu]\nonumber\\&\qquad - (\tau_{\mu+1}-\tau_\mu)
       		s_{ij}[\mu]a_{ij}[\mu], & &\forall i,j,\mu\label{u3}\\   	
			 &\sum_{i,j} a_{ij}[\mu]\leq m,&&\forall \mu\label{u6}\\
       &s_{min} \leq s_{ij}[\mu] \leq 1, &&\forall i,j,\mu\label{u7}\\ 
			 &0 \leq a_{ij}[\mu] \leq 1, &&\forall i,j,\mu\label{u8}
\end{align}
\end{subequations}
where $a_{ij}[\mu]\in \mathbb{R}$ is the fraction of  the time interval $[\tau_\mu,\tau_{\mu+1})$ for which job  $T_{ij}$ is executing on a processor at speed $s_{ij}[\mu]$ and~(\ref{u6}) specifies that the total workload in time interval $[\tau_\mu,\tau_{\mu+1})$ should be less than or equal to the system capacity. 
 

\begin{theorem}\label{thm2}  If a solution exists to~\eqref{ho1}--\eqref{ho7}, then a solution  to problem~\eqref{prob2d} exists. Furthermore, at least one valid schedule satisfying~\eqref{ho1}--\eqref{ho7}  can be  constructed from a solution to problem~\eqref{prob2d} with the same energy consumption.
\end{theorem}
\begin{IEEEproof}
A similar argument as Theorem 1 can be used to prove the existence of a solution. The simplest valid schedule can be constructed using McNaughton's wrap around algorithm~\cite{mc1959} for each time interval $[\tau_\mu,\tau_{\mu+1})$. During each interval $[\tau_\mu,\tau_{\mu+1})$, the taskset is schedulable since (i) the total density of the taskset does not exceed the total number of processors, which is guaranteed by contraint~(\ref{u6}), (ii) no task workload is greater than its deadline, which is ensured by constraint~(\ref{u8}) and (iii) task migration is allowed, which is our assumption. The optimal energy consumption relies on the fact that the total energy consumption does not depend on the order of which tasks are scheduled, but rather depends on the taskset's density, i.e.\ the objective value stays the same regardless of the number of discretization steps $M$ on the minor grid.
\end{IEEEproof}   

\subsection{Linear Program (LP-DVFS)}
Suppose now that the set of speed levels is finite, as is the case with a real processor. We denote with $s_q$ the  speed of the processor at level $q \in Q:=\{1,\ldots,l\}$, where $l$ is the total number of speed levels. 

Consider now the following finite-dimensional LP:
\begin{subequations} \label{probLP}
\begin{align}
& \underset{\begin{subarray}{c}
       x_{ij}[\cdot],a_{ij}^q[\cdot],\\
			 i\in I,j\in J_i,q\in Q\end{subarray}}
			 {\text{minimize}} \quad\mathrlap{\sum_{\mu,i,j,q}(\tau_{\mu+1}-\tau_\mu)  \ell(a_{ij}^q[\mu],s_q)}\\
	& \text{subject to } \nonumber\\
	     &x_{ij}[\Phi_A(T_{ij})]=\underline{x}_i & &\forall i,j \label{lp1}\\ 
			 &x_{ij}[\Phi_D(T_{ij})]= 0,  & &\forall i,j \label{lp2}\\ 
       &x_{ij}[\mu+1]=x_{ij}[\mu]\nonumber\\&\qquad - (\tau_{\mu+1}-\tau_\mu)\sum_{q}s_qa_{ij}^q[\mu], & &\forall i,j,\mu \label{lp3}\\   	
			 &\sum_{q}a_{ij}^q[\mu] \leq 1,&&\forall i,j,\mu \label{lp4}\\ 
			 &\sum_{i,j,q} a_{ij}^q[\mu]\leq m,&&\forall \mu\label{lp5}\\
			 &0 \leq a_{ij}^q[\mu] \leq 1,&&\forall i,j,q,\mu\label{lp6}
\end{align}
\end{subequations}
where
$a_{ij}^q[\mu]\in \mathbb{R}$ is the fraction of the time interval $[\tau_\mu,\tau_{\mu+1})$ for which job $T_{ij}$ is executing on a processor at speed $s_q$.

Constraint~(\ref{lp4}) assures that a task will not run on more than one processor at a time. Constraint~(\ref{lp5}) guarantees that a processor's workload will not exceed its capacity at a time. Lastly, constraint~(\ref{lp6}) provides upper and lower bounds on a fraction of a job's execution time at a specific speed level variable.

Note that, given a solution to the optimization problem~(\ref{probLP}), one could also employ McNaughton's wrap around algorithm~\cite{mc1959} to construct a valid schedule of processor assignments with the same energy consumption. In other words, Theorem~\ref{thm2} can be applied here as well to prove an existence of a valid schedule given a solution to problem~(\ref{probLP}).

\section{Simulation Results} \label{sec:sim}
\subsection{System, Processor and Task Models}
\begin{figure*}[t]
\centerline{
\subfloat[XScale]{\includegraphics[width=0.5\textwidth]{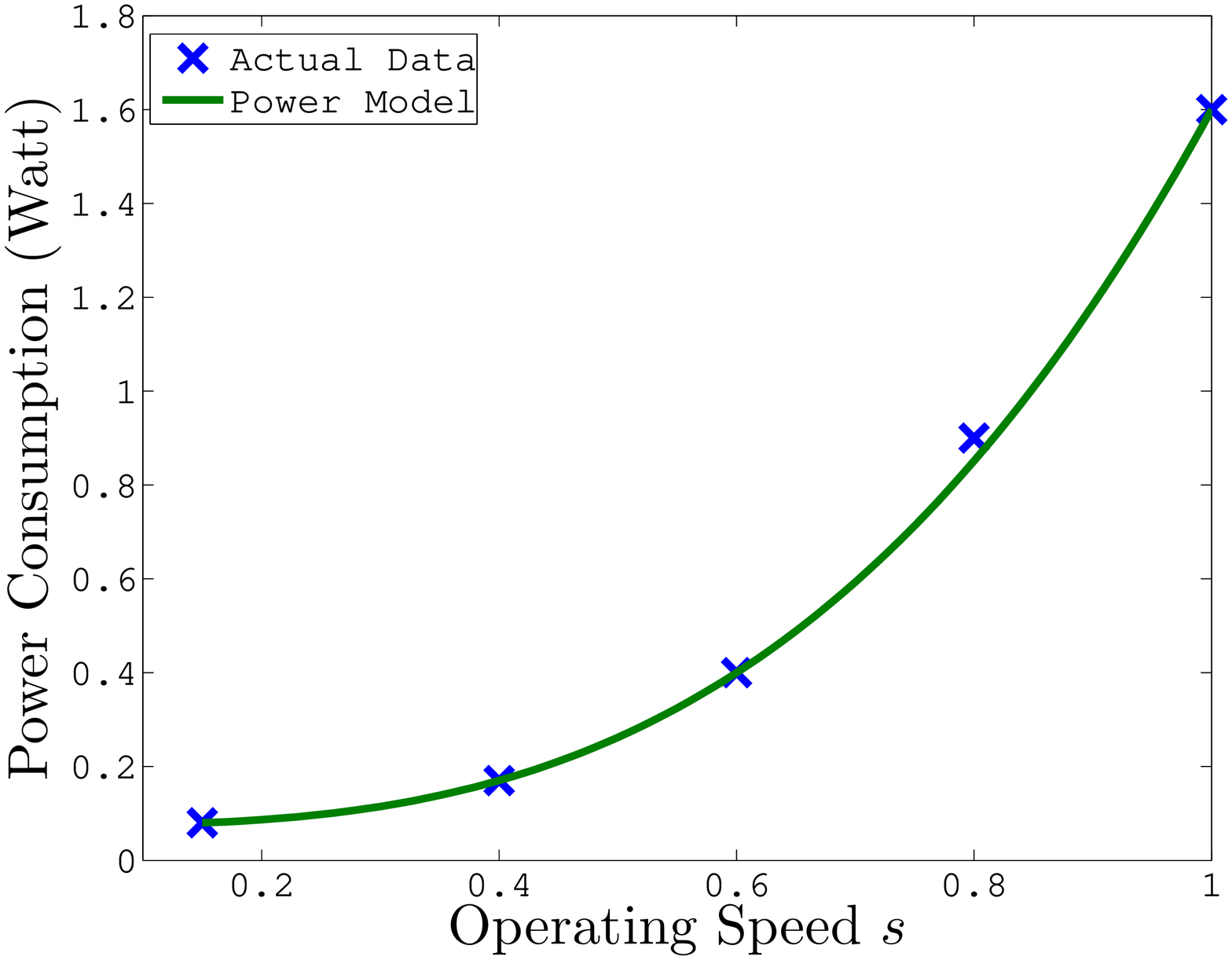}}
\subfloat[PowerPC 405LP]{\includegraphics[width=0.5\textwidth]{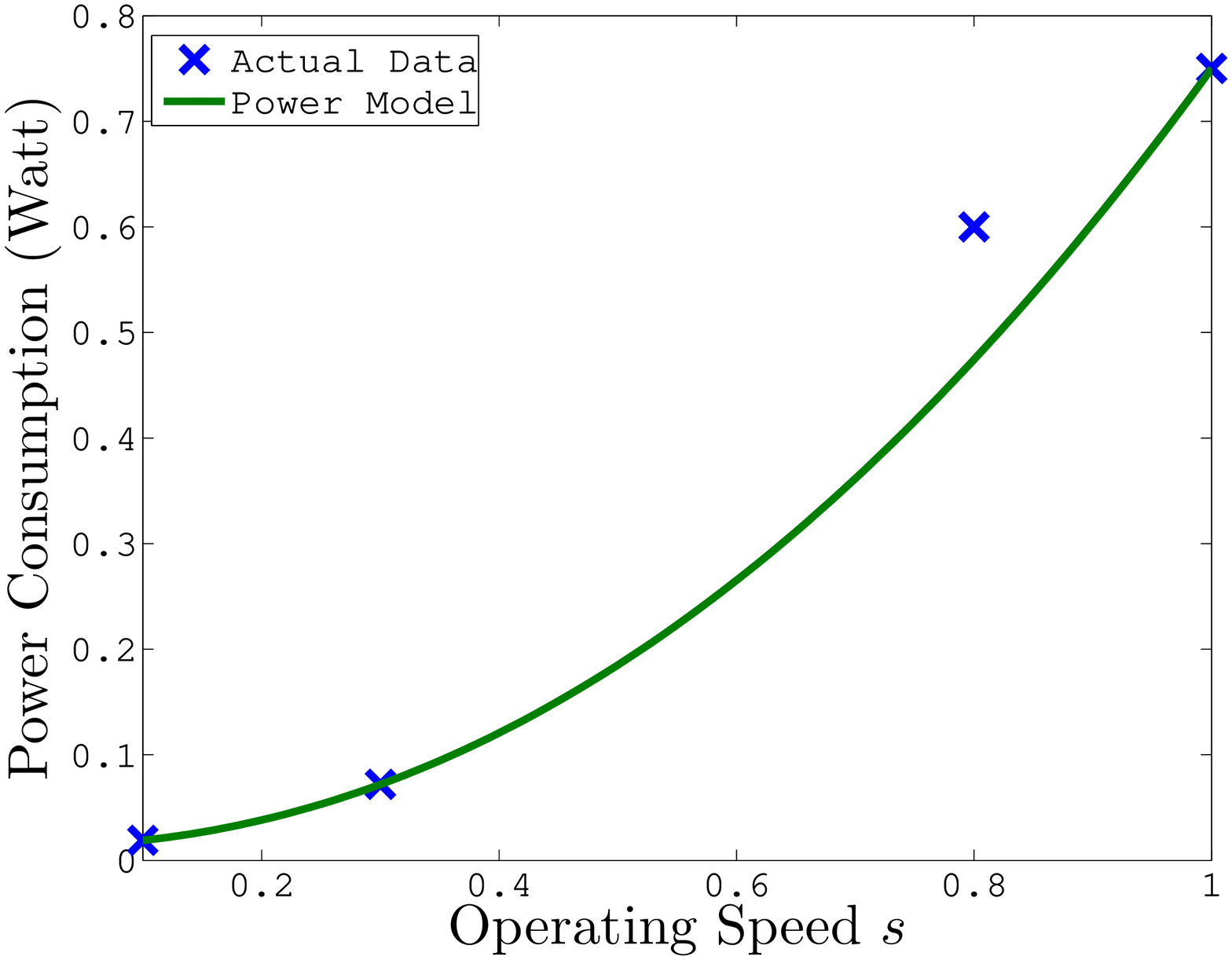}}} 
\caption{Actual Data versus Fitted Model}
\label{fig:fitplot}
\end{figure*}

The performance of solving the above optimization problems is evaluated on two models of practical commercial processors, namely an XScale and~PowerPC 405LP. The power consumption details of the two commercial processors, which have also been used in~\cite{Xu04, gang09, Wang12}, are given in Table~\ref{tab:pmsim}. The active power consumption models of the XScale and PowerPC~405LP shown in Table~\ref{tab:lsfit} were obtained by a polynomial curve fitting to the generic form~(\ref{eqn:Pd_lit}) (details are given in the appendix B). The plots of the actual data versus the fitted models are shown in Fig.~\ref{fig:fitplot}.

\begin{table*}[t]
\caption{Commercial Processor Details for Simulation} \label{tab:pmsim}
\centering
\begin{tabular}{|l|c|c|c|c|c|c|c|c|c|}\hline
Processor type & \multicolumn{5}{c}{XScale~\cite{xscale}} & \multicolumn{4}{|c|}{PowerPC 405LP~\cite{rusu04}} \\ \hline
Frequency (MHz) & 150 & 400 & 600 & 800 & 1000 & 33 & 100 & 266 & 333 \\
Speed	& 0.15 & 0.4 & 0.6 & 0.8 & 1.0 & 0.1 & 0.3 & 0.8 & 1.0 \\ 
Voltage (V) & 0.75 & 1.0 & 1.3 & 1.6 & 1.8 & 1.0 & 1.0 & 1.8 & 1.9 \\
Active Power (mW) & 80 & 170 & 400 & 900 & 1600 & 19 & 72 & 600 & 750 \\ \hline
Idle Power (mW) & \multicolumn{5}{c}{40~\cite{Xu04}} & \multicolumn{4}{|c|}{12} \\\hline
\end{tabular}
\end{table*}

\addtocounter{footnote}{1}

\begin{table*}[t]
\caption{Processor Power Consumption Fitted Model} \label{tab:lsfit}
\centering
\begin{tabular}{|l|l|c|}\hline
Processor type & \multicolumn{1}{c|}{Active Power Model} & MAPE$^{\decimal{footnote}}$ \\ \hline
XScale & $P_{active}(s):=1524.92s^{3.0269}+75.1092$ & 1.1236   \\ \hline
PowerPC 405LP & $P_{active}(s):=736.87s^{2.0990}+13.1333$ & 5.2323  \\ \hline
\end{tabular}
\end{table*}

\subsection{Comparison between Algorithms}
For a system with a continuous speed range, four algorithms were compared:~(i) MINLP-DVFS, (ii) NLP-DVFS, (iii) GP-SVFS, which represents a global energy/feasibility-optimal workload partitioning with constant frequency scaling scheme and (iv) GP-NoDVFS, which is a global workload allocation without frequency scaling scheme. For a system with discrete speed levels, three algorithms are compared: (i) LP-DVFS, (ii) GP-NoDVFS and (iii) GP-SDiscrete, which represents global energy/feasibility-optimal workload allocations with constant discrete frequency scaling. Note that the formulations of GP-SVFS and GP-SDiscrete are similar to~\cite{aydin2003,hoon2014}. Specifically, the GP-SVFS is based on a constant frequency scaling with global scheduling scheme, while~\cite{aydin2003} is a partitioning-based formulation and~\cite{hoon2014} is a generalized formulation for a two-type heterogeneous multi-core system. GP-SDiscrete is an extension to systems with discrete speed levels. Details on GP-SVFS, GP-NoDVFS and GP-SDiscrete are given below. 

GP-SVFS/GP-NoDVFS: Solve
\begin{subequations} \label{GSprob}
\begin{align}
& \underset{\begin{subarray}{c}
           s_k,k\in K \end{subarray}}{\text{minimize}}\mathrlap{\sum_{i,k}L\ell(u_{ik}(s_k),s_k)} \\
	     &\text{subject to }\sum_{k} u_{ik}(s_k) \leq 1,&&\forall i \label{gs1}\\
			 &\qquad\sum_{i}u_{ik}(s_k) \leq 1,&&\forall k \label{gs2}\\
			 &\qquad s_{min} \leq s_k \leq 1,&&\forall k~(GP-SVFS)\label{gs3}\\
			 &\qquad s_k = 1,&&\forall k~(GP-NoDVFS)\label{gs4}
\end{align}
\end{subequations}
where $u_{ik}(s_k)$ is the task density on processor $k$, i.e. $u_{ik}(s_k):=\underline{x}_i/(s_kf_{max}d_i$) and $s_k$ is the static execution speed of processor $k$. A task will not be executed on more than one processor at the same time due to~(\ref{gs1}) and the assigned workload will not exceed processor capacity due to~(\ref{gs2}). The difference between GP-SVFS and GP-NoDVFS lies in restriction on operating speed to be either a continous variable~(\ref{gs3}) or fixed at the maximum speed~(\ref{gs4}).

GP-SDiscrete: Determine a fraction of the workload $y_{ikq}$ of a task at a specific speed level and a processor speed level selection $z_{kq}$  such that:
\begin{subequations} \label{GSdisc}
\begin{align}
& \underset{\begin{subarray}{c}
           y_{ikq},z_{kq},i\in I,\\
	k\in K,q\in Q \end{subarray}}
	{\text{minimize}}	\
	 \mathrlap{\sum_{i,k,q} L\ell(u_{ikq}(y_{ikq})z_{kq},s_q)}\\
			 &\text{subject to }\sum_{k,q} y_{ikq}z_{kq} = 1,&&\qquad\forall i \label{gds1} \\
			 &\qquad\qquad \sum_{q} z_{kq} = 1,&&\qquad\forall k \label{gds2} \\
	     &\qquad\qquad \sum_{k,q} u_{ikq}(y_{ikq})z_{kq} \leq 1,&&\qquad\forall i \label{gds3}\\
			 &\qquad\qquad \sum_{i,q} u_{ikq}(y_{ikq})z_{kq} \leq 1,&&\qquad\forall k \label{gds4}\\
			 &\qquad\qquad 0 \leq y_{ikq} \leq 1,&&\qquad\forall i,k,q \label{gds5} \\ 
			 &\qquad\qquad z_{kq} \in \{0,1\},&&\qquad\forall k,q \label{gds6}
\end{align}
\end{subequations}
where $y_{ikq}$ represents a fraction of the workload of a task at a specific speed level and $z_{kq}$ is a speed level selection variable for processor~$k$, i.e.\ $z_{kq} = 1$ if a speed level $q$ is selected and $z_{kq} = 0$ otherwise. The constraints in (\ref{gds1}) guarantee that the total  workload of a task is equal to $1$. The constraints in~(\ref{gds2}) ensure that only one speed level is selected, (\ref{gds3}) assures that a task will be executed  on only one processor at a time, (\ref{gds5}) ensures that a processor workload capacity is not violated and~(\ref{gds6}) emphasises that the speed level selection variable is binary. Note that GP-SVFS and GP-NoDVFS are NLPs and GP-SDiscrete is an MINLP. 

\begin{table}[t]
\caption{Simulation Tasksets} \label{tab:taskset}
\centering
\begin{tabular}{|c|c|c|c|c|}\hline
$D$& $T_1$ & $T_2$ & $T_3$ & $T_4$\\ \hline
0.4 & (0.75,5,10) & (0.75,5,10) & (0.5,10,10) & (0.5,10,10) \\ \hline
0.6 & (1,5,10) & (1,5,10) & (1,10,10) & (1,10,10) \\ \hline
0.8 & (1.5,5,10) & (1.5,5,10) & (1,10,10) & (1,10,10) \\ \hline
1.0 & (2,5,10) & (2,5,10) & (1,10,10) & (1,10,10) \\ \hline
1.2 & (2.5,5,10) & (2.5,5,10) & (1,10,10) & (1,10,10) \\ \hline
1.4 & (3,5,10) & (3,5,10) & (1,10,10) &  (1,10,10)\\ \hline
1.6 & (3,5,10) & (3,5,10) & (2,10,10) & (2,10,10) \\ \hline
1.8 & (3.5,5,10) & (3.5,5,10) & (2,10,10) & (2,10,10) \\ \hline
2.0 & (4,5,10) & (4,5,10) & (2,10,10) & (2,10,10) \\ \hline
\multicolumn{5}{l}{Note: The first parameter of a task is $\underline{x}_i$; $c_i$ can be obtained }\\
\multicolumn{5}{l}{by multiplying $\underline{x}_i$ by $f_{max}$.} 
\end{tabular}
\end{table}


\begin{figure*}[t]
\centering
\subfloat[Xscale]{\label{2a}\includegraphics[width=0.5\textwidth]{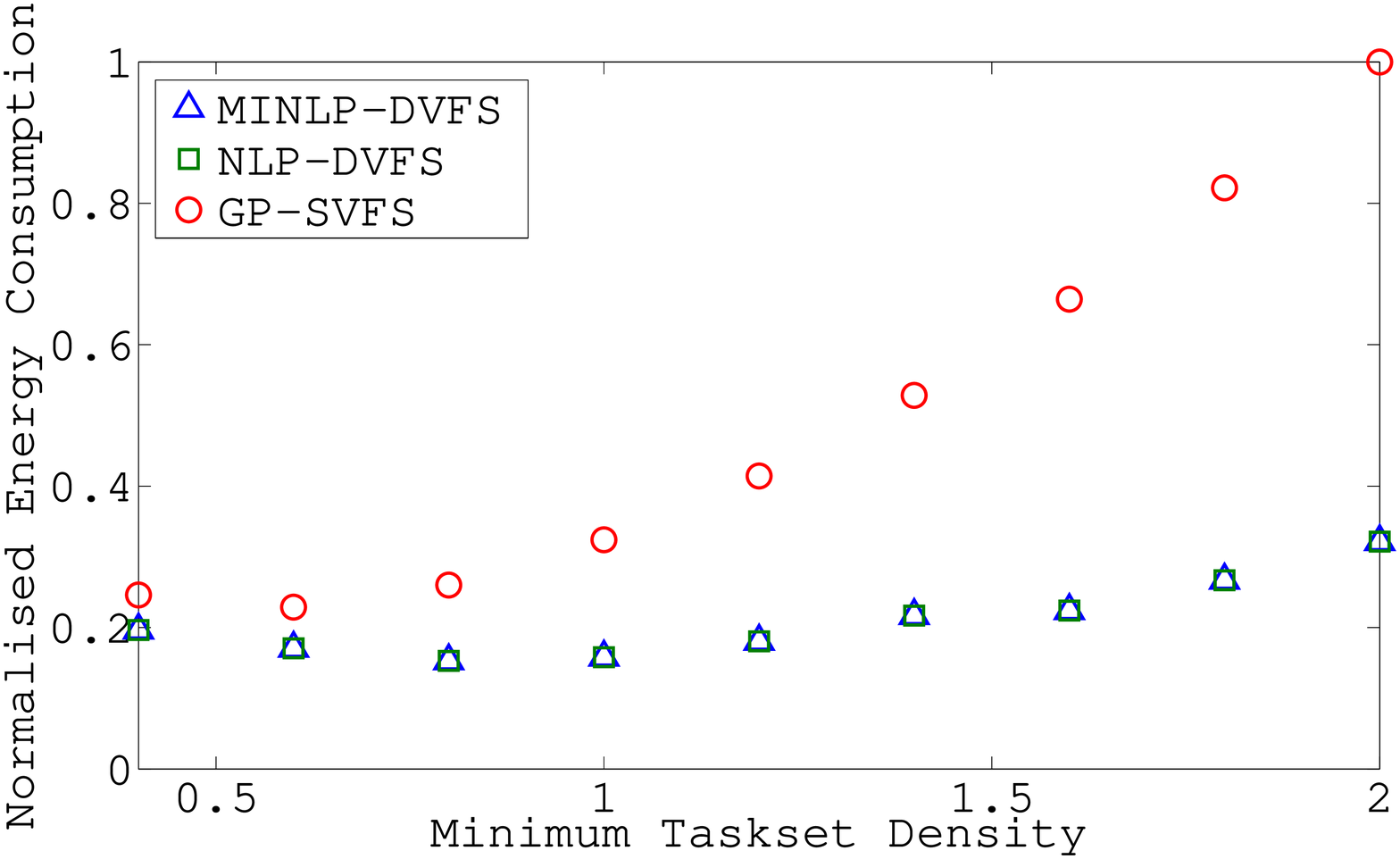}} 
\subfloat[PowerPC 405LP]{\label{2b}\includegraphics[width=0.5\textwidth]{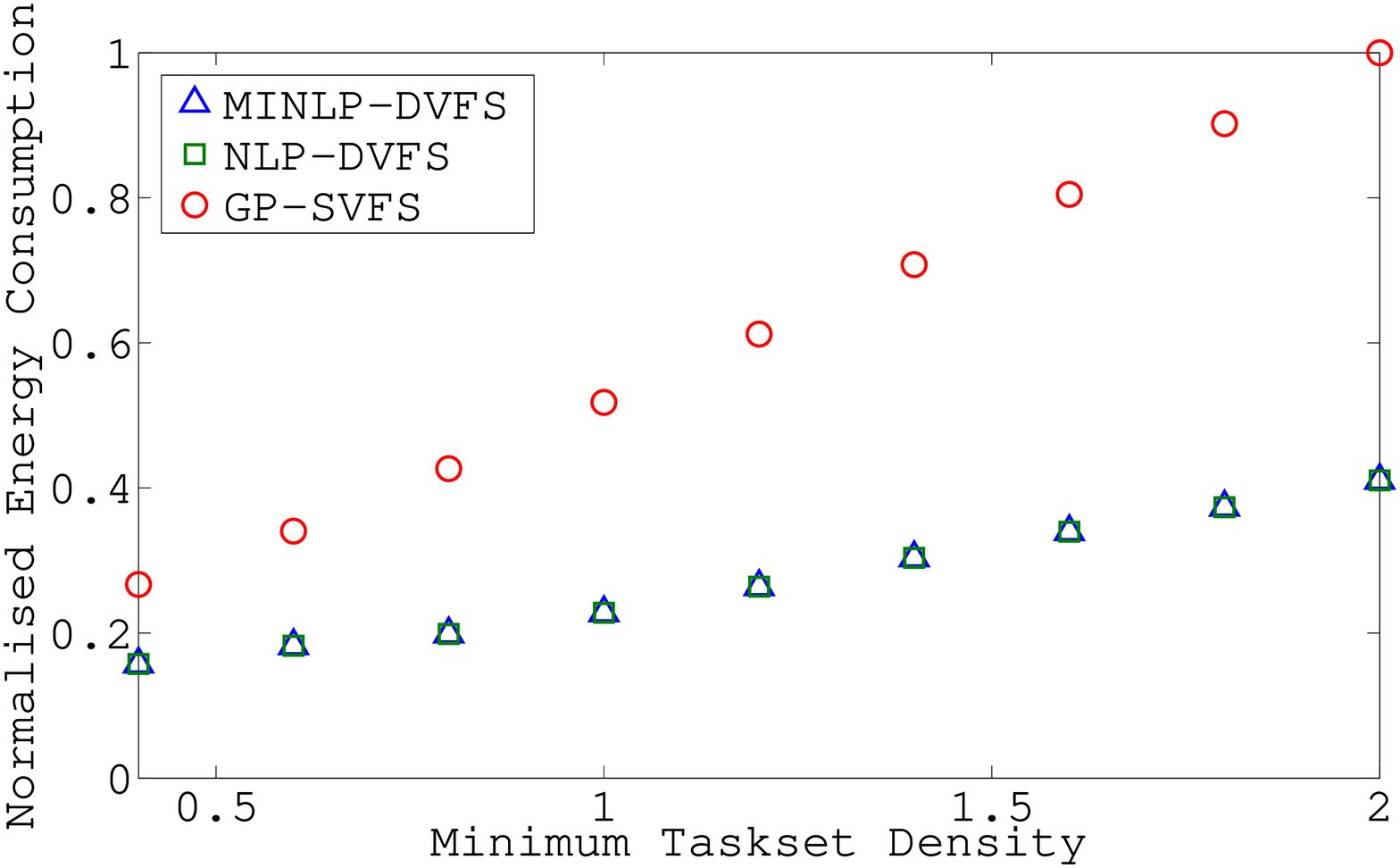}}
\caption{Simulation results for scheduling real-time tasks with constrained deadlines on two homogeneous processors with continuous speed range}
\label{fig:M2}
\end{figure*}

\begin{figure*}[t]
\centering
\subfloat[Xscale]{\label{3a}\includegraphics[width=0.5\textwidth]{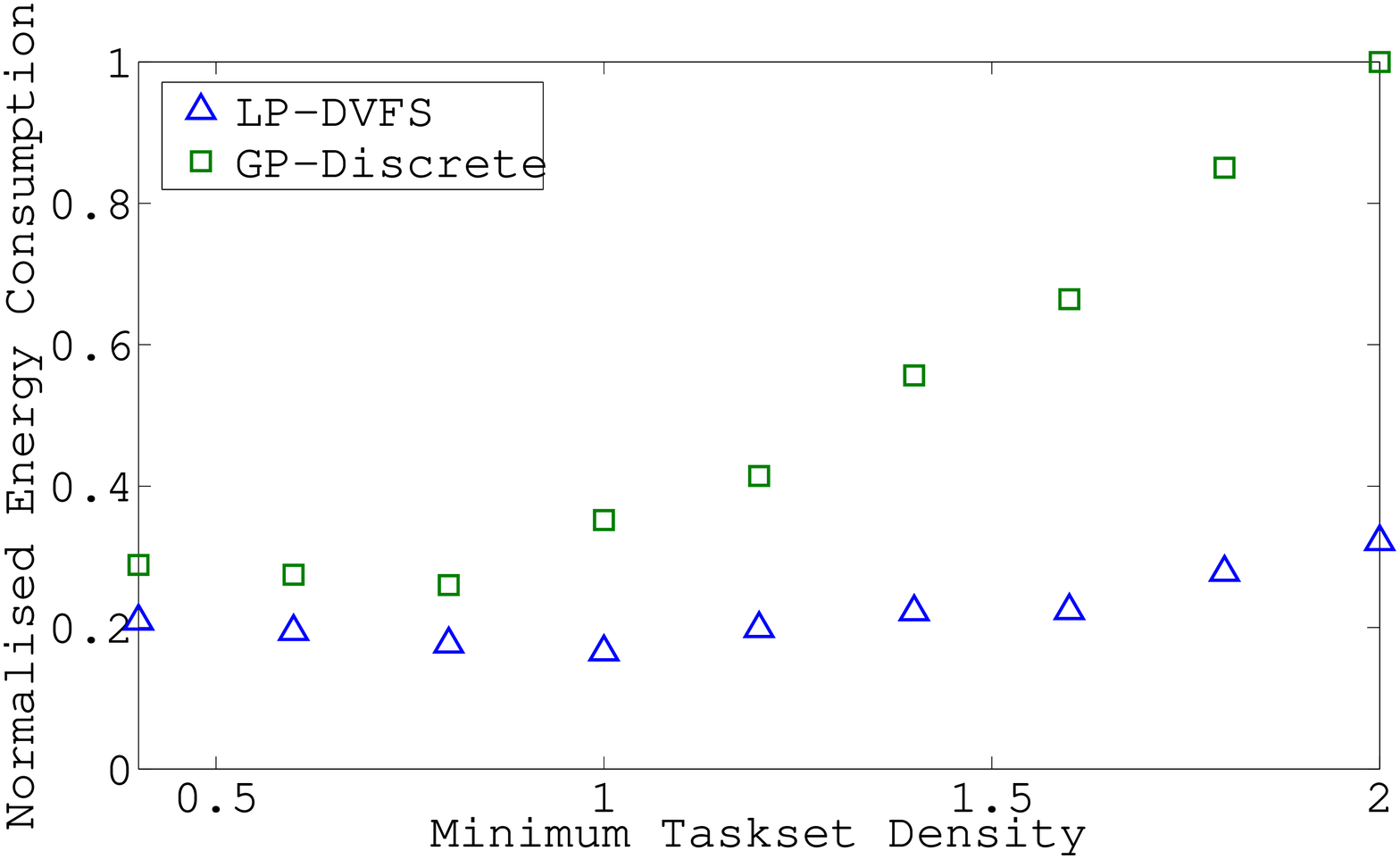}} 
\subfloat[PowerPC 405LP]{\label{3b}\includegraphics[width=0.5\textwidth]{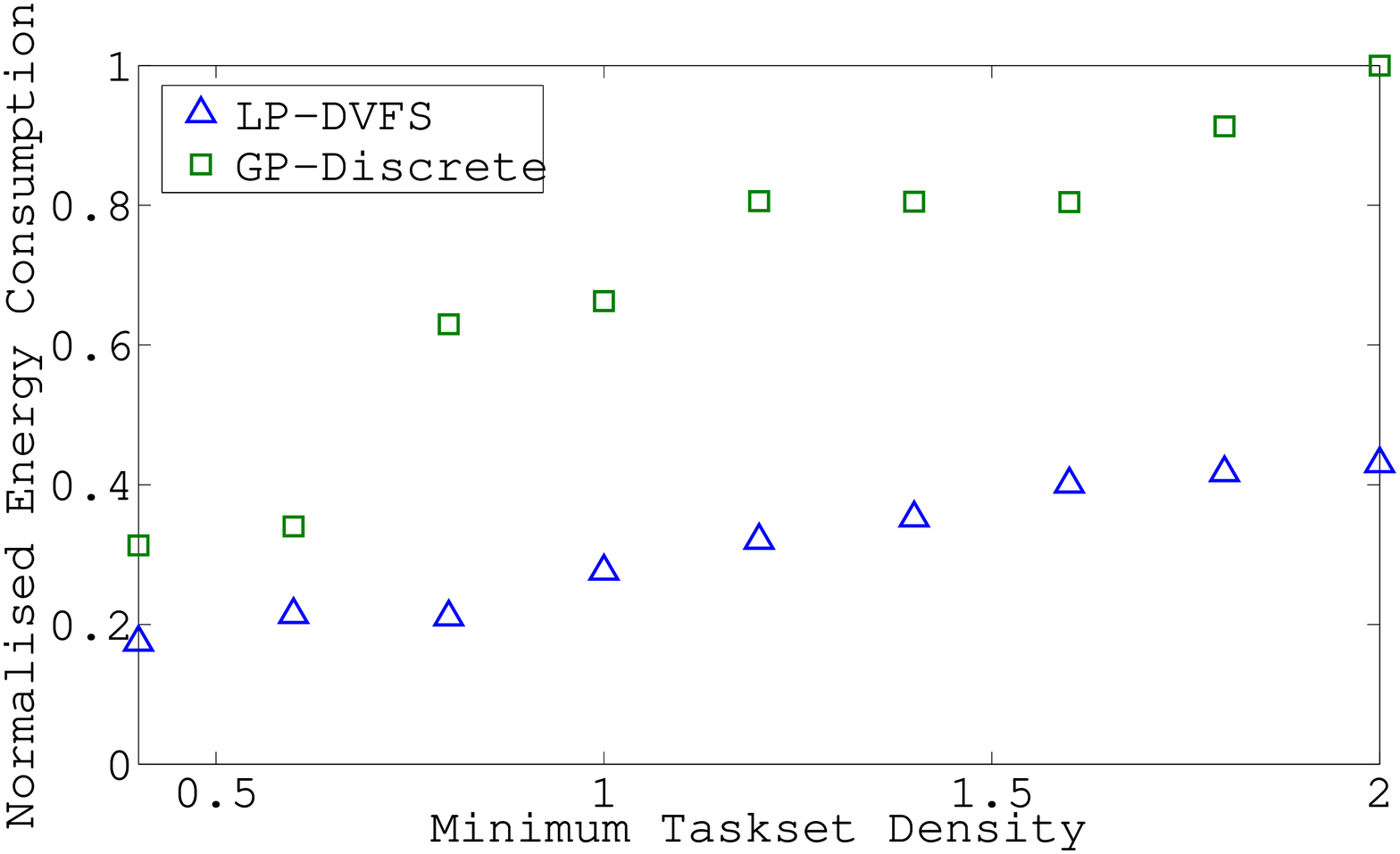}}
\caption{Simulation results for scheduling real-time tasks with constrained deadlines on two homogeneous processors with discrete speed levels}
\label{fig:M3}
\end{figure*}

\footnotetext[\value{footnote}]{See Appendix A.}

\subsection{Simulation Setup and Results}
For simplicity, we consider the case where four independent periodic real-time tasks with constrained deadlines need to be scheduled onto two homogeneous processors. The total energy consumption of each taskset in Table~\ref{tab:taskset} were evaluated. For the MINLP-DVFS, NLP-DVFS and LP-DVFS implementations, the major grid discretization step $N:=2$ because all tasksets for the simulation only have two distinct deadlines. For the MINLP-DVFS, we chose the minor grid discretization step $M:=4$, since there are at most four jobs in each major grid, which implies that there will be at most four preemptions within each major grid. All of our formulations were modelled using ZIMPL~\cite{Koch04} and solved by SCIP~\cite{Achterberg09}. 

Simulation results are shown in Figures~\ref{fig:M2} and~\ref{fig:M3}. The minimum taskset density $D$ is represented on the horizontal axis. The vertical axis represents the total energy consumption normalised by the energy used by GP-NoDVFS, i.e.\ less than~1 means that an algorithm does better than GP-NoDVFS. It can be seen from the plots that for a time-varying workload such as a constrained deadline taskset, our results from solving MINLP-DVFS, NLP-DVFS and LP-DVFS are energy optimal, while GP-SVFS, GP-Discrete and GP-NoDVFS are not. This is because our formulations are incorporated with time, which provides benefits on solving a scheduling problem with both time-varying workload (constrained deadline taskset) and constant workload (implicit deadline taskset). In general, compared to the constant speed profile, the time-varying speed profile saving increases as the minimum taskset density increases. It can also be noticed that the percentage saving of the Xscale is nonlinear compared to that of the PowerPC's. This is because the power consumption model of the Xscale is a cubic, while the PowerPC's power consumption model is a quadratic. However, it has to be mentioned that the energy saving percentage varies with the taskset, which implies that the number shown on the plots can be varied, but the significant outcomes stay the same. Lastly, the computation times to solve NLP-DVFS, LP-DVFS, GP-SVFS and  GP-SDiscrete are extremely fast, i.e.\  milliseconds using a general-purpose desktop PC with off-the-shelf optimization solvers, while the time to solve MINLP-DVFS can be up to an hour in some cases.

\section{Conclusions} \label{sec:concl}
Three mathematical optimization problems were proposed to solve a  homogeneous multiprocessor scheduling problem with a periodic real-time taskset. Though our MINLP and NLP formulations are both energy optimal and feasibility optimal, the computation time is high compared with heuristic algorithms. However, our LP formulation is computationally tractable and suitable for a system with discrete speed level set.  Moreover, we have shown via simulations that our formulations are able to solve a more general class of scheduling problem than existing work in the literature due to incorporating a scheduling dynamic model in the formulations as well as allowing for a time-varying executing speed profile. The simulation results illustrate the possibility that the time-varying speed profile can save energy up to 70\% compared to the constant speed profile. Possible future work could also include developing numerically efficient methods to solve the various mathematical optimization problems defined in this paper. One could also extend the ideas presented here to solve a dynamic scheduling problem with uncertainty in task execution time and to include slack reclamation for further energy reduction. 
  
\bibliographystyle{IEEEtran}
\bibliography{MSAshort}

\appendices
\section{Mean Absolute Percentage Error (MAPE)}
\begin{equation}
\text{MAPE}:=\frac{1}{k}\sum_{i=1}^k\frac{|F(z(i))-y(i)|}{|y(i)|}\times 100,
\end{equation}
\noindent where $\frac{|F(z(i))-y(i)|}{|y(i)|}$ is the magnitude of the relative error in the $i^{th}$ measurement, $z\mapsto F(z)$ is the estimated function, $z$ is the input data, $y$ is the actual data and $k$ is the total number of fitted points.

\section{Curve Fitting}
The following shows the curve fitting formulation to obtain the active power consumption function of the XScale and PowerPC 405LP with the objective of minimizing the MAPE: 
\begin{subequations} 
\begin{align}
&\underset{\alpha,\beta,P_s}{\text{minimize}}~100\times\frac{1}{\gamma}\sum_{i=1}^\gamma |\alpha s_i^\beta+P_s-P_{active}(s_i)|/P_{active}(s_i) \nonumber\\
	& \text{subject to } \nonumber\\
	& \qquad\qquad P_s \geq 0 \label{a1}\\
	& \qquad\qquad \alpha \geq 0 \\
	& \qquad\qquad \beta \geq 1
\end{align}
\end{subequations}

\noindent where $\alpha,~\beta$ and $P_s$ are parameters to be fitted, $\gamma$ is the total number of speed levels of a processor. The speed level of a processor $s_i$ and the active power consumption of a processor $P_{active}(s_i)$ are measurement data shown in Table~\ref{tab:pmsim}.  

\end{document}